\documentclass[12pt]{JHEP3}
\usepackage{mathrsfs}
\usepackage{amsmath,amssymb}
\usepackage{epsfig}
\usepackage{relsize}
\usepackage{bbm} 			
\usepackage{dcolumn}  			

\newcommand{\beq}{\begin{equation}}
\newcommand{\eeq}{\end{equation}}
\newcommand{\be}{\begin{equation}}
\newcommand{\ee}{\end{equation}}
\newcommand{\bea}{\begin{eqnarray}}
\newcommand{\eea}{\end{eqnarray}}
\newcommand{\ba}{\begin{array}}
\newcommand{\ea}{\end{array}}

\newcommand{\bra}[1]{\langle{#1}|}
\newcommand{\ket}[1]{|{#1}\rangle}

\newcommand{\nn}{\nonumber}

\newcommand{\ns} \normalsize

\author{Per Sundin \\ \\
 {\it Humboldt-Universit\"at zu
Berlin, Institut f\"ur Physik,\\ Newtonstra{\ss}e 15, D-12489
Berlin, Germany}\\ \email{per.sundin@physik.hu-berlin.de}}
\abstract{We perform a detailed study of bosonic type IIA string theory in a large light-cone momentum / near plane wave limit of AdS$_4 \times \mathbbm{CP}_3$. In order to attain this we derive the Hamiltonian up to cubic and quartic order in number of fields and calculate the energies for string excitations in a R$ \times$S$^2 \times $S$^2$ subspace. The computation for the string energies is performed for arbitrary length excitations utilizing an unitary transformation which allows us to remove the cubic terms in the Hamiltonian. We then rewrite a recent set of proposed all loop Bethe equations in a light-cone language and compare their predictions with the obtained string energies. We find perfect agreement.  \vspace{20pt} \\
   \\

}

\title{The AdS$_4 \times \mathbbm{CP_3}_3$ string and its Bethe equations in the near plane wave limit}
\preprint{HU-EP-08/58}
\begin{document}
\newpage
\section{Introduction}
It has been a long standing belief that the low energy dynamics of multiple M2 branes in M-theory can effectively be described by a three dimensional gauge theory \cite{Schwarz:2004yj}. Recently, a precise duality was suggested by Aharony, Bergman, Jafferis and Maldacena \cite{Aharony:2008ug} where they proposed a new exciting $AdS / CFT$ correspondence relating M-theory on AdS$_4 \times$S$^7/\mathbbm{Z}_k$ with three dimensional $\mathcal{N}=6$, $SU(N) \times SU(N)$ Chern Simons theory\footnote{For a nice review, see \cite{klose}.}.

Combining the level $k$ of the Chern Simons theory with the rank $N$ of the gauge group, one can introduce a 't Hooft coupling as $\lambda=N/k$. For small values of the coupling, it was shown in \cite{Minahan:2008hf} that the SU(4) R symmetry sector of the gauge theory \cite{Benna:2008zy} can be mapped to an integrable spin chain. Taking $\lambda$ to be large, M-theory on AdS$_4 \times$S$^7/\mathbbm{Z}_k$ can effectively be described by a type IIA string theory on AdS$_4 \times \mathbbm{CP}_3$ \cite{Aharony:2008ug} which leaves us with a new weak / strong coupling duality between a boundary gauge theory and a ten dimensional string theory ($AdS_4 / CFT_3$). Since a lot is known from the original $AdS_5 / CFT_4$ correspondence, there has been a remarkable progress in understanding both the gauge theory and the string theory side of the $AdS_4 / CFT_3$ duality \cite{Bak:2008cp,Bak:2008vd,Nishioka:2008gz,Gaiotto:2008cg,Gromov:2008qe,Gromov:2008fy,Gromov:2008bz,McLoughlin:2008he,McLoughlin:2008ms,Alday:2008ut,Ahn:2008hj,Kreuzer:2008vd}.

Even though the rapid developments, there are still a lot of things to be learned. Most pressing is the question about quantum integrability on both sides of the correspondence. On the gauge theory side, integrability has been demonstrated to hold at leading order in perturbation theory \cite{Minahan:2008hf}, while the dual string theory is integrable at the classical level\footnote{Or, to be precise, the dual string theory formulated as a coset model on $OSP(2,2|6)/SO(1,3)\times U(3)$ with 24 fermions is classically integrable \cite{Arutyunov:2008if}. Classical integrability for the full type IIA model with 32 fermions remains to be proved. We thank the authors in \cite{Gomis:2008jt} for valuable comments regarding this point.} \cite{Arutyunov:2008if,Stefanski:2008ik}.\\
Although the quantum regime of the string theory has been probed by various string configurations in \cite{Nishioka:2008gz,Gaiotto:2008cg,McLoughlin:2008he,McLoughlin:2008ms,Alday:2008ut,Krishnan:2008zs,Kristjansen:2008ib},  it is nevertheless safe to say that string quantum integrability remains largely unknown.

In the present paper we hope to shed some light on the question of quantum string integrability by performing a detailed study of the bosonic string in a near plane wave limit. For similar limits in the $AdS_5 / CFT_4$ duality, see \cite{Berenstein:2003gb,Callan:2004uv,Frolov:2006cc} and references therein.

Some aspects of the bosonic near plane wave AdS$_4 \times \mathbbm{CP}_3$ string  have been extracted in \cite{Astolfi:2008ji} where the authors calculated energy corrections for string states in a SU(2)$\times$SU(2) subsector using second order perturbation theory and $\zeta$-function regularization. These energies were compared perturbatively with the predictions from a set of conjectured all loop Bethe equations \cite{Gromov:2008qe}. Even though a nice result, we feel there is a need to make an even more careful analysis than done in \cite{Astolfi:2008ji}. Most pressing is the question about factorized scattering which is a crucial ingredient for any integrable field theory. While \cite{Astolfi:2008ji} found agreement with the Bethe equations in \cite{Gromov:2008qe}, it was only established for operators built out of two oscillators. A stronger test of integrability would be to calculate the energy shifts for an arbitrary number of oscillators which would allow for precise statements about the factorization properties of the excitations to be made.

We also feel there is need for a deeper understanding of the string Lagrangian and Hamiltonian. As was noticed in \cite{Astolfi:2008ji}, a novel feature of the type IIA AdS$_4 \times \mathbbm{CP}_3$ string is that it exhibits cubic interactions. We show that these can be analyzed through a set of successive canonical transformations. This allows us to reformulate the cubic interactions in terms of additional quartic terms with the advantage that first order perturbation theory can be applied for calculating string energies. This is important since it gives a finite answer when calculating the energies and we do not have to use any $\zeta$-function regularization schemes. It is also important since second order perturbation theory includes summations over intermediate states and it is not immediately clear why one can ignore fermionic contributions as done in \cite{Astolfi:2008ji}.

We will perform our investigations in a uniform light-cone gauge \cite{Arutyunov:2005hd, Frolov:2006cc} which gives a very convenient expansion scheme for the near plane wave limit. The gauge also allow us to rewrite the all loop Bethe equations \cite{Gromov:2008qe} in a compact set of uniform light-cone gauge Bethe equations (ULCB). The simple form of the ULCB equations allows for an analytical comparison with the results obtained from the string Hamiltonian.

The paper is organized as follows; In section two we discuss some general properties of the string Lagrangian and its Hamiltonian. We then introduce light-cone coordinates $x^\pm=\frac{1}{2}(\phi\pm t)$, where $t$ is a global time coordinate and $\phi$ is an angle in $\mathbbm{CP}_3$, and impose the uniform light cone gauge $x^+=\tau$ together with $P_+$ constant (where $P_+$ is the conjugate momenta of $x^-$) \cite{Arutyunov:2005hd,Frolov:2006cc}. The near plane wave limit is equivalent to a large $P_+$ limit, and following \cite{Frolov:2006cc}, we expand the string Hamiltonian up to quartic order in fields. The section is concluded by showing that the point particle dynamics are fully captured by the quadratic Hamiltonian.

In section three we perform a perturbative quantization and calculate the energy shifts for arbitrary numbers of operators in the SU(2)$\times$SU(2) subsector of the theory. This subsector describes strings within a $R\times S^2 \times S^2$ subspace of the AdS$_4 \times \mathbbm{CP}_3$ background. To calculate the energy shifts we remove the cubic terms in the Hamiltonian through a unitary transformation.

In section four we rewrite the proposed all loop Bethe equations of \cite{Gromov:2008qe} in a uniform light-cone basis \cite{Hentschel:2007xn}. We solve the ULCB equations analytically for the SU(2)$\times$ SU(2) sector and find perfect agreement with the energies obtained from the string theory computation.

We end the paper with a brief discussion and outlook together with several appendices.
\section{Bosonic type IIA string on AdS$_4 \times \mathbbm{CP_3}_3$}
The main interest of this paper is to perform a detailed study of the bosonic string propagating on an AdS$_4 \times \mathbbm{CP_3}_3$ background. The natural starting point will be the string Lagrangian,
\bea
\label{L}
\mathscr{L}=-\frac{1}{2}\, \sqrt{\widetilde{\lambda}} \, \gamma^{\alpha \beta} \, G_{MN} \,\partial_\alpha \,x^M \,\partial_\beta \,x^N,
\eea
where $\gamma^{\alpha \beta}$ is the Weyl invariant worldsheet metric, with  $det\, \gamma =-1$. Throughout the paper we will use Greek / Latin indices for worldsheet / space-time quantities. As done in \cite{McLoughlin:2008he}, we define a modified 't Hooft coupling, $\widetilde{\lambda}$, given by
\bea
\label{coupling}
\widetilde{\lambda}=2\pi^2 \,\lambda,
\eea
to emphasize the close resemblance to the $AdS_5 / CFT_4$ case. The string length, $\sigma$, is chosen to take values between $[0,2\pi]$.

The space-time metric, $G_{MN}$, is defined through
\bea
ds^2=ds^2_{AdS_4}+4\,ds^2_{CP_3},
\eea
where we use the coordinates \cite{Arutyunov:2008if,Callan:2004uv}
\bea
ds^2_{CP^3}= \frac{d\bar{\omega}_i d \omega_i}{1+|\omega |^2}-\frac{d \bar{\omega}_j \omega_j \bar{\omega}_i d \omega_i}{(1+|\omega|^2)^2}, \quad
ds^2_{AdS_4} = - \Big(\frac{1+\frac{z_a^2}{4}}{1-\frac{z_a^2}{4}}\Big)^2dt^2+\frac{1}{(1-\frac{z_a^2}{4})^2}dz_a^2,
\eea
with complex $\omega_i$, and $i,j,a,b\in \{1,2,3\}$. Furthermore, the complex coordinates are parameterized as \cite{Arutyunov:2008if},
\bea
\label{parametrization}
\omega_3=(1-y)e^{i \phi},\quad \omega_2=\frac{1}{\sqrt{2}}r_2 e^{\frac{i}{2} \phi},\quad \omega_1=\frac{1}{\sqrt{2}}r_1 e^{\frac{i}{2} \phi},
\eea
with real $y$ and complex $r_1, r_2$. The two complex coordinates parameterize the two S$^2$ in $\mathbbm{CP}_3$. Later, we will study excitations within this subspace.

Using (\ref{parametrization}) gives
\bea
\label{metric}
&& \frac{1}{2}\,ds^2_{CP^3}= \\ \nn
&& \frac{1}{1+|\omega|^2}\Big\{d\phi^2\Big((1-y)^2+\frac{1}{8}\bar{r} \cdot r - \frac{(1-y)^4+\frac{1}{2}(1-y)^2 \bar{r} \cdot r +\frac{1}{16}(\bar{r} \cdot r)^2}{1+|\omega|^2}\Big) \\ \nn
&& + dy^2\Big(1-\frac{(1-y)^2}{1+|\omega|^2}\Big)+\frac{1}{2} d \bar{r}_s dr_t \Big(\delta_{st}-\frac{1}{2}\, \frac{ r_s \bar{r}_t}{1+|\omega|^2}\Big)+\frac{1}{2}dy(1-y)\frac{d \bar{r} \cdot r + \bar{r} \cdot dr}{1+|\omega|^2}\Big) \\ \nn
&& + \frac{i}{4}d\phi(d\bar{r} \cdot r-\bar{r} \cdot dr)\Big(1-2 \frac{(1-y)^2}{1+|\omega|^2}-\frac{1}{2} \, \frac{\bar{r} \cdot r}{1+|\omega|^2}\Big)\Big\},
\eea
After imposing a suitable gauge, this will be the coordinates we expand the Lagrangian (\ref{L}) in.

As is normally done, we will also combine $t$ and $\phi$ into a light-cone pair as
\bea
x^\pm = \frac{1}{2}(\phi \pm t).
\eea
The theory is invariant under global shifts in the two $x^\pm$ coordinates, where the associated conserved Noether charges are
\bea
\label{Ppm}
P_\pm = \pm \Delta + J,
\eea
with
\bea
\Delta = -\frac{1}{2 \pi} \int_0^{2\pi} \,d\sigma \, p_t, \qquad J = \frac{1}{2 \pi} \int_0^{2\pi} \,d\sigma \, p_\phi.
\eea
Here $\Delta$ measures the space-time energy with respect to the AdS time $t$ and $J$ denotes the conserved angular momentum for the angular coordinate $\phi$. Note that the transverse coordinates $r_1,r_2,z_a$ and $y$ are not charged under neither $\Delta$ or $J$.
\subsection{Hamiltonian dynamics}
For the upcoming analysis, it is very convenient to rewrite (\ref{L}) in first order form,
\bea
\label{firstorderL}
\mathscr{L}=p_M\, \dot{x}^M-\mathcal{H}.
\eea
Due to the two dimensional diffeomorphic invariance on the worldsheet, the Hamiltonian is just a sum of constraints \cite{Callan:2004uv}
\bea
\label{fullH}
\mathcal{H}=\frac{1}{2 \gamma^{00}}\big(p_M p_N G^{MN}+\widetilde{\lambda} \,x'^M x'^N G_{MN}\big)-\frac{\gamma^{01}}{\gamma^{00}}\, x'^M \, p_M,
\eea
where the prime denotes derivatives with respect to the length parameter of the string. The equation of motion for the worldsheet metric gives
\bea
\label{constraint}
C_1: \qquad p_M p_N G^{MN}+\widetilde{\lambda} \,x'^M x'^N G_{MN}=0, \quad C_2: \qquad   x'^M \, p_M=0.
\eea
The first constraint, $C_1$, will be solved perturbatively for the light-cone Hamiltonian and solving the second constraint, $C_2$, allow us to express $x'^-$ in terms of transverse fields. Integrating this equation gives the level matching condition which should be imposed on physical states.

However, before solving the constraints, we need to impose a suitable gauge. In this paper we will employ a uniform light-cone gauge \cite{Arutyunov:2005hd,Frolov:2006cc},
\bea
\label{gauge}
x^+=\tau, \qquad P_+ = \textrm{ Constant. }
\eea
In this gauge\footnote{One consequence of this gauge is that the string length $r$ goes like $r \sim P_+ / \sqrt{\widetilde{\lambda}}$. Depending on the scalings of the coupling and the light-cone momenta, the string length may or may not be finite \cite{Arutyunov:2006gs}. For the problem at hand however, we can rescale $\sigma$ so that it takes values between $[0,2\pi]$.}, (\ref{firstorderL}) becomes
\bea
\mathscr{L}=P_-+p_m\, \dot{x}^m,
\eea
where $m$ labels the transverse coordinates. Thus, the light-cone Hamiltonian is given by\footnote{We will suppress the $LC$ subscript in the subsequent discussion.}
\bea
\label{Hconst}
\mathcal{H}_{L.C}=-P_-=\Delta-J.
\eea
Using (\ref{constraint}) we will solve this equation perturbatively for $P_-$.
\subsection{Large $P_+$ expansion}
We will do a perturbative study in a near plane wave limit defined by \cite{Arutyunov:2005hd}
\bea
\label{exp}
P_+ \Rightarrow \infty, \qquad \widetilde{\lambda} \sim P_+^2,
\eea
together with the following rescalings of the transverse coordinates
\bea
\label{scalingsandcoupling}
P_M \Rightarrow \sqrt{\frac{P_+}{2}} \, P_M \qquad x^M \Rightarrow \sqrt{\frac{2}{P_+}}x^M, \qquad \widetilde{\lambda}'=\frac{4\, \widetilde{\lambda}}{P_+^2},
\eea
where we have defined the effective coupling $\widetilde{\lambda}'$ which remains finite in the large $P_+$ limit. This is similar but not identical to the effective coupling $\lambda'=\lambda/J^2$, which is kept fix in the usual large $J$ limit \cite{Berenstein:2003gb}.

As was discussed in \cite{Frolov:2006cc}, the expansion scheme above is equivalent to an expansion in number of fields. Thus, (\ref{Hconst}) has an expansion as\footnote{$\mathcal{H}_3 \sim \mathcal{O}(P_+^{-1/2})$ and $\mathcal{H}_4 \sim \mathcal{O}(P_+^{-1})$.}
\bea
\label{Hexp}
\mathcal{H}=\mathcal{H}_2+\mathcal{H}_3+\mathcal{H}_4+...,
\eea
where the subscript denotes the number of fields in each expansion term. The presence of a cubic interaction term is a novel feature compared to the well known AdS$_5 \times$S$^5$ case \cite{Astolfi:2008ji}.

Expanding the first constraint in (\ref{constraint}) to quadratic order, gives
\bea
\label{H2}
\mathcal{H}_2=\frac{1}{2 }\Big(p_{y}^2+p_i^2+p_a^2+y^2+z_a^2+\frac{1}{4}x_i^2+\widetilde{\lambda}'\big(y'^2+z_a'^2+x_i^2\big)\Big).
\eea
where we have expressed $r_1$ and $r_2$ in terms of four real coordinates $x_i$,  with $i \in \{1,2,3,4\}$. As was first observed in \cite{Gaiotto:2008cg,Nishioka:2008gz}, one of the $\mathbbm{CP}_3$ coordinates, $y$, combines on the same mass level as the AdS$_4$ coordinates $x_i$. This seem to occur only at the quadratic level and we will see later that the higher order terms in the Hamiltonian separates $y$ from the other AdS$_4$ coordinates\footnote{However, it could be that this combination of coordinates occurs again if one identify a proper canonical transformation to push the cubic interactions up to quartic order. This transformation is somewhat complicated to find due to the presence of derivative terms in $\mathcal{H}_2$ and $\mathcal{H}_3$.}.
	
The next to leading order term in (\ref{Hexp}) only has dependence on the $\mathbbm{CP}_3$ fields and is given by\footnote{We have simplified the expression using that up to a total derivative,
$4 \widetilde{\lambda}(y' \,x_i \,x_i'+y\,(x')^2)=-4\widetilde{\lambda}y \, x_i\, x_i''$.}
\bea
\label{H3}
&& \mathcal{H}_3=\\ \nn
&& \frac{1}{4 \sqrt{2\, P_+}}\Big(x_i^2+4\,y^2-4 \,\widetilde{\lambda}' \,x_i\, x_i''+8\,\widetilde{\lambda}'\,y'^2+4\big(x_2 \, p_1-x_1\, p_2+x_4\, p_3-x_3\,p_4\big) \\ \nn
&& -4\,p_i^2-8\,p_y^2\Big)y -\frac{1}{\sqrt{2\,P_+}}x_i\,p_i \, p_y,
\eea
The quartic Hamiltonian is quite complicated and to simplify the notation, we split it up into three separate parts
\bea
\mathcal{H}_4=\mathcal{H}_4^{AdS_4}+\mathcal{H}_4^{AdS_4/CP_3}+\mathcal{H}_4^{CP_3}.
\eea
The pure AdS$_4$ part is simply
\bea
\mathcal{H}_4^{AdS_4}=\frac{\widetilde{\lambda}'}{P_+}\,\big(z_1^2+z_2^2+z_3^2\big)\big(z_1'^2+z_2'^2+z_3'^2\big)
\eea
and the term with both AdS$_4$ and $\mathbbm{CP}_3$ dependence is given by
\bea
&& \mathcal{H}_4^{AdS_4/CP_3}=\\ \nn
&& \frac{1}{2\,P_+}\Big\{\widetilde{\lambda}'\Big(z_a^2\,(x_i'^2+y'^2)-z_a'^2\,(y^2+\frac{1}{4}\,x_i^2)\Big)
+z_a^2\,(p_i^2+p_y^2)-p_a^2\,(y^2+\frac{1}{4}\,x_i^2)\Big\}.
\eea
The more complicated $\mathbbm{CP}_3$ contribution is
\bea
\label{H4}
&& \mathcal{H}_4^{CP_3}=\\ \nn
&& \frac{1}{32 \, P_+}\Big\{4\,x_i^2\,y^2-(x_i^2)^2+24\,y^4+20\,x_i^2\, p_i^2+  12\big(x_1^2\,p_2^2+x_2^2\,p_1^2+x_3^2\,p_4^2+x_4^2\, p_3^2\big) \\ \nn
&& +4 \big((x_1^2+x_2^2)(p_3^2+p_4^2)+(x_3^2+x_4^2)(p_1^2+p_2^2)\big)+4\,x_i^2\,p_y^2+48\,y^2\,p_y^2 \\ \nn
&& +16\,y^2(x_2\,p_1-x_1\,p_2+x_4\,p_3-x_3\,p_4)+16\big((2\,x_1\,x_4-x_2\,x_3)p_1\,p_4+(2\,x_2\,x_3-x_1\,x_4)p_2\,p_3 \\ \nn
&& +(2\,x_2\,x_4+x_1\,x_3)p_2\,p_4+(2\,x_1\,x_3+x_2\,x_4)p_1\,p_3+x_1\,x_2\,p_1\,p_2+x_3\,x_4\,p_3\,p_4\big)+64\,y\,p_y\,x_i\,p_i\\ \nn
&& -4\widetilde{\lambda}'\Big(2\big((x_1^2+x_2^2)(x_1'^2+x_2'^2)+(x_3^2+x_4^2)(x_3'^2+x_4'^2)\big)+3\, x_i^2\,x_i'^2-8\, y \,y'\,x_i\,x_i' \\ \nn
&& +4\big((x_2\,x_3-x_1\,x_4)(x_2'\,x_3'-x_1'\,x_4')+(x_1\,x_3+x_2\,x_4)(x_1'\,x_3'+x_2'\,x_4')\big)+x_i^2\,y'^2-12\,y^2\,y'^2\Big)\Big\}.
\eea
A nice thing with the coordinates we use is that the Hamiltonian does not have any $x^-$ dependence. At the order we are interested in, this coordinate simply drops out of the equations. Therefore, the only effect of the $C_2$ constraint in (\ref{constraint}) is the level matching condition. Nevertheless,
as can be seen, the Hamiltonian (\ref{Hexp}) is still considerably more complicated than the AdS$_5 \times$ S$^5$ one in \cite{Frolov:2006cc}.
\subsection{Point particle limit}
Before we proceed with a detailed study of the Hamiltonian we need to resolve one issue. We expect that the point particle dynamics should be fully governed by the quadratic Hamiltonian. However, upon taking the point particle limit, $\sigma \rightarrow 0$, of (\ref{Hexp}) we see that there are both cubic and quartic non derivative terms that survives. We denote these $\mathcal{H}_3^0$ and $\mathcal{H}_4^0$ and their explicit form can be found in (\ref{zeromodepiece}) and (\ref{zeromodepiece1}).

These terms can be removed by performing successive canonical transformations on the Hamiltonian (\ref{Hexp}). We start by recalling how a generating functional $V(x,p)$ acts on a general phase space function $f(x,p)$
\bea
&& f(x,p) \Rightarrow \\ \nn
&& f(x,p)+\{V(x,p),f(x,p)\}_{P.B}+\frac{1}{2!} \,\{V(x,p),\{V(x,p),f(x,p\}_{P.B}\}_{P.B}+...
\eea
The generating functional we are about to construct will be perturbative in $P_+$,
\bea
\label{V}
V=V_3+V_4,
\eea
where $V_3$ is of order $P_+^{-1/2}$ and $V_4$ is of order $P_+^{-1}$. The leading order part, $V_3$, is constructed so that it removes the cubic terms. Thus, $V_3$, has the property
\bea
\nn
\{V_3, \mathcal{H}^0_2\}_{P.B}=-\mathcal{H}_3^0+\mathcal{O}(P_+^{-1}),
\eea
where the full expression for $V_3$ can be found in (\ref{V3}). At order $P_+^{-1}$, this term will induce additional quartic terms through
\bea
\label{quarticaddition}
&& \mathcal{O}(P_+^{-1}): \qquad \mathcal{H}^0_{Add}=\{V_3,\mathcal{H}_3^0\}_{P.B}+\frac{1}{2}\,\{V_3,\{V_3,\mathcal{H}_2^0\}_{P.B}\}_{P.B}+\mathcal{O}(P_+^{-3/2})\\ \nn
&& = \frac{1}{2}\,\{V_3,\mathcal{H}_3^0\}_{P.B}+\mathcal{O}(P_+^{-3/2}).
\eea
This additional term is simpler than $\mathcal{H}_4^0$ in (\ref{zeromodepiece1}), but nevertheless quite involved, see (\ref{quarticaddition2}).

We construct the next to leading order term in (\ref{V}) so that it remove the original and additional quartic parts of the Hamiltonian,
\bea
\label{V4prop}
\{V_4, \mathcal{H}_0\}_{P.B}=-\frac{1}{2}\, \{ V_3,\mathcal{H}_0\}-\mathcal{H}_4^0.
\eea
The explicit expression for $V_4$ can be found in the appendix, equation (\ref{V4}).

With this we have constructed a generating functional $V$ with the desired property
\bea
\{V,\mathcal{H}^0\}_{P.B}=-\mathcal{H}_3^0-\mathcal{H}_4^0+\mathcal{O}(P_+^{-3/2}).
\eea
It is important to note that this does not imply that we can neglect the non derivative terms for the case of non-zero $\sigma$. The generating functional becomes significantly more complicated since it involves non-local effects through terms like
\bea
\nn
\frac{\delta \mathcal{H}_2}{\delta X}\sim \frac{\delta \mathcal{H}_2}{\delta X}+\frac{\delta (\partial_\sigma X)}{\delta X}\cdot\frac{\delta \mathcal{H}_2}{\delta (\partial_\sigma X)}.
\eea
It is still plausible though that one can remove all the non derivative terms through a non local canonical transformation which will add additional derivative quartic terms. However, for the problem at hand this will not be necessary.
\section{Field expansion, unitary transformations and energy shifts}
We now have the full Hamiltonian to quartic order and are in position to investigate the detailed consequences of it. One of the aims with the present work is to do a perturbative calculation of the energy shift in closed subsectors of the theory. To do that we will follow the well known procedure of expanding the coordinates in Fourier modes, promoting oscillators to operators through the quantization process and calculating the energy shifts in perturbation theory. Except for the novel presence of cubic terms, and the complication arising from that, the chapter that follows will share many similarities with \cite{Frolov:2006cc}.
\subsection{Field expansions and quantization}
We start with expanding the coordinates in fourier modes,
\bea
\label{modexpansion}
&& z_a=i \sum_k e^{-ik\sigma}\frac{1}{\sqrt{2\,\Omega_k}}\big(\hat{z}_{a,k}-\hat{z}_{a,-k}^\dagger\big), \quad  p_a= \sum_k e^{-ik\sigma}\sqrt{\frac{\Omega_k}{2}}\big(\hat{z}_{a,k}+\hat{z}_{a,-k}^\dagger\big), \\ \nn
&& x_i=i \sum_k e^{-ik\sigma}\frac{1}{\sqrt{2\,\omega_k}}\big(x_{i,k}-x_{i,-k}^\dagger\big), \quad  p_i= \sum_k e^{-ik\sigma}\sqrt{\frac{\omega_k}{2}}\big(x_{i,k}+x_{i,-k}^\dagger\big), \\ \nn
&&
y= i \sum_k e^{-ik\sigma}\frac{1}{\sqrt{2\,\Omega_k}}\big(y_{k}-y_{-k}^\dagger\big), \quad p_y= \sum_k e^{-ik\sigma}\sqrt{\frac{\Omega_k}{2}}\big(y_{k}+y_{-k}^\dagger\big),
\eea
where the frequencies are given by
\bea
\omega_k=\sqrt{\frac{1}{4}+\widetilde{\lambda}'\,k^2}, \qquad \Omega_k=\sqrt{1+\widetilde{\lambda}'\,k^2}.
\eea
The Fourier coefficients are promoted to operators through usual commutation relations,
\bea
\label{commutationrelations}
[\hat{z}_{a,k},\hat{z}_{b,l}^\dagger]=\delta_{ab}\,\delta_{kl}, \quad [x_{i,k},x_{j,l}^\dagger]=\delta_{ij}\, \delta_{kl}, \quad [y_k,y_l^\dagger]=\delta_{kl}.
\eea
Using the mode expansions (\ref{modexpansion}), the free Hamiltonian becomes
\bea
\mathcal{H}_2=\sum_{k}\Big( \omega_k \, x^\dagger_{i,k}\, x_{i,k}+\Omega_k\big(y^\dagger_k \, y_k + \hat{z}^\dagger_{a,k}\,\hat{z}_{a,k}\big)\Big),
\eea
and the second constraint in (\ref{constraint}) equals
\bea
\mathcal{V}= \sum_{k}\, k \,\big(  x^\dagger_{i,k}\, x_{i,k}+y^\dagger_k \, y_k + \hat{z}^\dagger_{a,k}\,\hat{z}_{a,k}\big),
\eea
where physical states has to satisfy $\mathcal{V}\, \ket{Phys}=0$.
\subsection{Removing cubic terms}
There are several ways to obtain the energy shifts of physical states \cite{Klose:2006zd,Frolov:2006cc,Astolfi:2008ji}. The most straightforward way is to calculate them using perturbation theory. Since we have the cubic interactions, it seems that we have to resort to second order perturbation theory. This would complicate things quite drastically. Not only would the calculation be more involved, but we would have to sum over intermediate, zeroth order, eigenstates. In principle this should also include the fermionic eigenstates, which we do not include in this analysis\footnote{Nevertheless, it seems to work when restricting to closed subsectors, see \cite{Astolfi:2008ji}.}.
However, armed with the experience from the previous section, we could try to transform the cubic part away and then calculate energy shifts using only first order perturbation theory. Since we are now including stringy effects, performing a canonical transformation directly on the coordinates is quite complicated. Instead we will construct the equivalent transformation on the level of oscillators\footnote{This section closely follow the construction outlined in \cite{Frolov:2006cc}.}. The oscillator picture is simpler since a coordinate and its derivative is, up to a mode number dependent factor, almost the same.

Performing the transformation on the quantum level, the construction of $V$ is such that
\bea
e^{i V}\mathcal{H}e^{-i V}=-\mathcal{H}_3+\mathcal{O}(P_+^{-1}).
\eea
$V$ is cubic in oscillators and has a general form
\bea
V=V^{+++}+V^{++-} + h.c,
\eea
where the superscript denotes the number of creation and annihilation operators. The explicit construction of $V$ in terms of components of $\mathcal{H}_3$ is straightforward. We start by writing $\mathcal{H}_3=G^{+++}+G^{++-}+h.c$, with
\bea
\label{H3split}
G^{+++}=\sum_{\substack{k,l,m \\a,b,c}} G_{a,b,c;k,l,m}^{+++}\,X^{\dagger,a}_k\,X^{\dagger,b}_l\,X^{\dagger,c}_m, \quad G^{++-}=\sum_{\substack{k,l,m \\a,b,c}} G_{a,b,c;k,l,m}^{++-}\,X^{\dagger,a}_k\,X^{\dagger,b}_l\,X^c_m,
\eea
where $a,b,c$ and $k,l,m$ are space-time / mode number indices and the set $a,b,c$ can denote any kind of oscillator, $\hat{z},x$ or $y$. The components of $V$ can now directly be constructed from (\ref{H3split}) \cite{Frolov:2006cc},
\bea
\label{VpppVppm}
&& V^{+++}=-i \sum_{\substack{k,l,m \\a,b,c}} \,\frac{G^{+++}_{a,b,c;k,l,m}}{\omega_{a,k}+\omega_{b,l}+\omega_{c,m}}
\,X^{\dagger,a}_k\,X^{\dagger,b}_l\,X^{\dagger,c}_m , \\ \nn
&& V^{++-}=-i \sum_{\substack{k,l,m \\a,b,c}} \,\frac{G^{++-}_{a,b,c;k,l,m}}{\omega_{a,k}+\omega_{b,l}-\omega_{c,m}}
\,X^{\dagger,a}_k\,X^{\dagger,b}_l\,X^{c}_m,
\eea
where $\omega_{a,k}$ is either $\omega_k$ or $\Omega_k$ depending on the index $a$. The explicit form of $V^{+++}$ and $V^{++-}$ is presented in (\ref{Vunitaryrepresentation}) and (\ref{Vunitaryrepresentation1}).

With (\ref{VpppVppm}) we have by construction that
\bea
\label{construction}
i[V,\mathcal{H}_2]=-\mathcal{H}_3,
\eea
and as in the point particle analysis, this transformation will induce additional quartic terms through
\bea
\label{Hadd}
\mathcal{O}(P_+^{-1}): \qquad \mathcal{H}_{Add}=i[V,\mathcal{H}_3]-\frac{1}{2}\{V^2,\mathcal{H}_2\}+V\,\mathcal{H}_2\,V=\frac{i}{2}[V,\mathcal{H}_3].
\eea
We do not try to add any additional higher order terms to $V$ to simplify the quartic terms. Adding counter terms is quite simple when dealing with transformations on the level of the coordinates, but doing it with oscillators complicates things. This does not really matter anyway since the cubic terms in the Hamiltonian (\ref{Hexp}) do contribute to the physical spectrum. While we might be able to simplify things, we can not expect to remove these terms completely.

Before we end this section, let us make some comments on the normal ordering of the Hamiltonian. We can probably take the original cubic and quartic Hamiltonian to be normal ordered \cite{Astolfi:2008ji}. However, this implies that the quartic additional contribution, coming from the unitary transformation, will be subject to normal ordering ambiguities. Basically we will get a quadratic normal ordering contribution of the form $C_{a,b,m,n} \, X_{m}^{\dagger,a} \, X_{n}^b$.

For the energy shift we will calculate, these terms can be shown to vanish upon $\zeta$ -function regularization. This is a consequence of the fact that when we evaluate a specific matrix element, the term above will always leave a sum over at least one internal index. Very schematically we will have something as, $\sum_k \widetilde{C}_k$, where $\widetilde{C}_k$ is a function of mode numbers and the coupling $\widetilde{\lambda}'$. Performing a perturbative expansion in the coupling gives an expansion in positive powers of mode numbers. Each term in this expansion can be shown to vanish due to the $\zeta$ -function identity \cite{Motl:1995cd}
\bea
\nn
\sum_{m\in \mathbbm{Z}} (m+\alpha)^s=0,
\eea
where $\alpha$ is a constant and $s>0$.
\subsection{Energy shifts for SU(2)$\times$SU(2) states}
Having removed the quartic terms, we can resort to first order perturbation theory to calculate the energy shifts from the Hamiltonian (\ref{Hexp}). Due to the complexity, we will focus on a subsector $R \times S^2 \times S^2$, which is spanned by the transverse coordinates $x_i$. To make the U(1) charges of each S$^2$ manifest, we do a complex redefinition of the $x_{i,k}$ oscillators as follows,
\bea
\label{redef}
&& x_{1,k}\Rightarrow \frac{1}{\sqrt{2}}\,\big(z_{1,k}+\widetilde{z}_{1,k}\big), \qquad x_{2,k} \Rightarrow \frac{i}{\sqrt{2}}\,\big(\widetilde{z}_{1,k}-z_{1,k}\big), \\ \nn
&& x_{3,k}\Rightarrow \frac{1}{\sqrt{2}}\,\big(z_{2,k}+\widetilde{z}_{2,k}\big), \qquad x_{4,k} \Rightarrow \frac{i}{\sqrt{2}}\,\big(\widetilde{z}_{2,k}-z_{2,k}\big).
\eea
The upshot of this transformation is that each oscillator is distinctly charged under the U(1)'s, as can be seen in Table \ref{tab:}.
\begin{table}[t]
\centering
\begin{tabular}{|c|c|c|c|c|} \hline
 &$z_{1,k}$ & $\widetilde{z}_{1,k}$ &$z_{2,k}$ & $\widetilde{z}_{2,k}$ \\ \hline
U(1)& 1 & -1 & 0 & 0 \\
U($\bar{1}$) &0 & 0 & 1 & -1\\ \hline
\end{tabular}
\caption{Charge table for complex oscillators}
\label{tab:}
\end{table}
For the gauge theory Bethe equations, the sector we want to match with the string theory predictions consist of the operators $A_{i_1}$ and $B_{j_1}$ transforming under the (1/2,0) and (0,1/2) of SU(2)$\times$SU(2) \cite{Minahan:2008hf}. The string states that correspond to these operators are the oscillators $\{z_{1,k}, \widetilde{z}_{2,k}\}$. Thus, the states we will calculate the energy shifts for are
\bea
\label{instate}
\ket{m_M,...,m_1,\bar{n}_{\bar{N}},...,1}=z_{1,m_M}^\dagger \,...\,z_{1,m_1}^\dagger\, \widetilde{z}_{2,\bar{n}_{\bar{N}}}^\dagger \, ... \, \widetilde{z}_{2,\bar{n}_{1}}^\dagger \, \ket{0},
\eea
for arbitrary numbers of oscillators $M$ and $\bar{N}$. For simplicity we will consider distinct mode numbers only. The explicit calculation for the energy shifts of the above states is straightforward but somewhat tedious. To make the calculation easier to follow, we will focus on the original quartic Hamiltonian (\ref{H4}) and the additional quartic contribution (\ref{Hadd}) separately.

The contributing part for the original quartic Hamiltonian is given by putting all AdS$_4$ excitations and the $y$ excitation to zero and performing the limit (\ref{redef}). Using (\ref{H4orgOscillator}), we find that
\bea
\label{shift1}
&& \bra{\bar{n}_1,...,\bar{n}_{\bar{N}} \, m_1,...,m_M}\,(\mathcal{H}_4^{CP_3}\big)\,\ket{m_M,...,m_1,\bar{n}_{\bar{N}},...,\bar{n}_1}= \\ \nn
&& -\frac{1}{4\, P_+}\Big\{\sum_{i=1}^M \sum_{j=1}^{\bar{N}}\frac{(m_i-\bar{n}_j)^2\,\widetilde{\lambda}'+2\, \omega_{m_i}\, \omega_{\bar{n}_j}}{\omega_{m_i}\, \omega_{\bar{n}_j}}\Big\} \\ \nn
&& -\frac{1}{16\, P_+}\Big\{\sum_{\substack{i,j \\ i\neq j}}^{\bar{N}} \frac{1+5\,(\bar{n}_i+\bar{n}_j)^2\widetilde{\lambda}'-4\big(\omega_{\bar{n}_i}^2+\omega_{\bar{n}_i} \, \omega_{\bar{n}_j}+\omega_{\bar{n}_j}^2\big)}{\omega_{\bar{n}_i}\, \omega_{\bar{n}_j}} \\ \nn
&& +\sum_{\substack{i,j \\ i \neq j}}^M \frac{1+5\,(m_i+m_j)^2\widetilde{\lambda}'-4\big(\omega_{m_i}^2+\omega_{m_i} \, \omega_{m_j}+\omega_{m_j}^2\big)}{\omega_{m_i}\, \omega_{m_j}}\Big\}.
\eea
For the additional terms coming from the unitary transformation (\ref{Hadd}), the calculation is a bit more involved. Using (\ref{HaddOsc1}) and (\ref{HaddOsc2}) in the appendix, gives
\bea
\label{shift2}
&& \bra{\bar{n}_1,...,\bar{n}_{\bar{N}} \, m_1,...,m_M}\,(\mathcal{H}_{Add}\big)\,\ket{m_M,...,m_1,\bar{n}_{\bar{N}},...,\bar{n}_1}= \\ \nn
&& \frac{\bar{N}\,M}{2\,P_+}-\frac{1}{16\,P_+}\Big\{\sum_{\substack{i,j \\ i\neq j}}^{\bar{N}}\frac{\Omega_{\bar{n}_i+\bar{n}_j}^2+4\,\omega_{\bar{n}_i}\,\omega_{\bar{n}_j}}{\omega_{\bar{n}_i}\,\omega_{\bar{n}_j}}
+\sum_{\substack{i,j \\ i\neq j}}^M\frac{\Omega_{m_i+m_j}^2+4\,\omega_{m_i}\,\omega_{m_j}}{\omega_{m_i}\,\omega_{m_j}}\Big\}.
\eea
By adding these two terms together we obtain that the energy shift for the SU(2)$\times$SU(2) sector is given by
\bea
\label{finalE}
&& \Delta E^{su2 \times su2}= \frac{\bar{N}\,M}{2\,P_+} -\frac{1}{4\, P_+}\Big\{\sum_{i=1}^M \sum_{j=1}^{\bar{N}}\frac{(m_i-\bar{n}_j)^2\,\widetilde{\lambda}'+2\, \omega_{m_i}\, \omega_{\bar{n}_j}}{\omega_{m_i}\, \omega_{\bar{n}_j}}\Big\} \\ \nn
&& -\frac{1}{16\, P_+}\Big\{\sum_{\substack{i,j \\ i \neq j}}^{\bar{N}} \frac{6\,\Omega_{\bar{n}_i+\bar{n}_j}^2-4\big(1+\omega_{\bar{n}_i}^2+\omega_{\bar{n}_j}^2\big)}{\omega_{\bar{n}_i}\, \omega_{\bar{n}_j}} +\sum_{\substack{i,j \\ i \neq j}}^M \frac{6\,\Omega_{m_i+m_j}^2-4\big(1+\omega_{m_i}^2+\omega_{m_j}^2\big)}{\omega_{m_i}\, \omega_{m_j}}\Big\}.
\eea
This is one of the main results of this paper. For two excitations, and in a different coordinate system, the corresponding energy shift were calculated in \cite{Astolfi:2008ji}. The result we obtain here holds for general number of impurities and is of a much simpler structural form. The simplicity is a consequence of the uniform light-cone gauge. This gauge choice also exhibit similar simplifications in the AdS$_5 \times$ S$^5$ case \cite{Frolov:2006cc}.

In the next section we will show that the energy shift (\ref{finalE}) is exactly reproduced by the Bethe equations of \cite{Gromov:2008qe} in a light-cone basis.
\section{Large $P_+$ expansion of the all loop asymptotic Bethe equations}
As has been known a long time, the dilatation operator of $\mathcal{N}=4$ SYM can be mapped to a spin chain Hamiltonian \cite{Minahan:2002ve,Beisert:2005fw,Beisert:2006ez,Spill:2008tp,Spill:2008yr}. This line of research, initiated by Minahan and Zarembo in \cite{Minahan:2002ve}, led to an enormous progress in understanding the exact spectrum of operators on both sides of the $AdS_5 / CFT_4$ correspondence. Astoundingly, it seems that much of what has been learned in the original duality can be repeated for the $AdS_4 / CFT_3$ correspondence. For example, to leading order the dilatation operator of the Chern Simons theory was demonstrated to be equivalent to a SU(4) spin chain Hamiltonian \cite{Minahan:2008hf}. Soon after, this was followed by an all loop proposal in \cite{Gromov:2008qe}.

In the section below we will match the energy shifts obtained from diagonalization of the string Hamiltonian with predictions from the all loop Bethe equations of \cite{Gromov:2008qe} written in a light-cone basis.
\subsection{Light-cone Bethe equations}
We start by writing down the all loop Bethe equations \cite{Gromov:2008qe} for the reduced  SU(2)$\times$SU(2) sector
\bea
\label{bethe}
&&\Big(\frac{x^+(p_k)}{x^-(p_k)}\Big)^L=\prod_{k\neq j}^{M}\,S(p_k,p_j)\,\prod_{j=1}^{M}\sigma(p_k,p_j)\,\prod_{j=1}^{\bar{N}}\sigma(p_k,q_j) \\ \nn
&& \Big(\frac{x^+(q_k)}{x^-(q_k)}\Big)^L=\prod_{k\neq j}^{\bar{N}}\,S(q_k,q_j)\,\prod_{j=1}^{\bar{N}}\sigma(q_k,p_j)\,\prod_{j=1}^{M}\sigma(q_k,p_j),
\eea
where the S-matrix is given by
\bea
S(p_k,p_j)=\frac{\Phi(p_k)-\Phi(p_j)+i}{\Phi(p_k)-\Phi(p_j)-i},
\eea
with
\bea
\Phi(p_k)=\cot\,\frac{p_k}{2}\sqrt{\frac{1}{4}+4\,h(\lambda)^2\, \sin^2\frac{p_k}{2}}.
\eea
The rapidities, $p_k$ and $q_k$, has to satisfy the momentum constraint
\bea
\label{momentumconstraint}
\sum_{i=1}^M p_k + \sum_{j=1}^{\bar{N}} q_j=0.
\eea
The function $h(\lambda)$ interpolates between $\lambda$ for small values of the 't Hooft coupling and $\sqrt{\lambda/2}$ for large values \cite{Gaiotto:2008cg,Nishioka:2008gz}. The variables $x^\pm$ and $\Phi$ are related through
\bea
x^\pm + \frac{1}{x^\pm}=\frac{1}{h(\lambda)}\Big(\Phi \pm \frac{i}{2}\Big).
\eea
For the dressing phase, we will only need the leading order part \cite{Arutyunov:2004vx} which can be written in terms of conserved charges as
\bea
&& \sigma(p_k,p_j)= \\ \nn
&& \exp\{2i\,\sum_{r=0}^\infty \Big(\frac{h(\lambda)^2}{4}\Big)^{r+2}\big(Q_{r+2}(p_k)\,Q_{r+3}(p_j)-Q_{r+3}(p_k)\,Q_{r+2}(p_j)\big)\Big\},
\eea
where the charges $Q_{r}(p_k)$ are given by
\bea
Q_r(p_k)=\frac{2\,\sin(\frac{r-1}{2}\,p_k)}{r-1}\Big(\frac{\sqrt{\frac{1}{4}+4\,h(\lambda)^2\sin^2\frac{p_k}{2}}-\frac{1}{2}}{h(\lambda)^2\sin\frac{p_k}{2}}\Big)^{r-1}.
\eea
The light-cone energy can be expressed through the dispersion relation
\bea
\label{betheE}
\Delta-J=\sum_{j=1}^{M}\Big(\sqrt{\frac{1}{4}+4\,h(\lambda)^2\sin^2\frac{p_j}{2}}-\frac{1}{2}\Big)
+\sum_{j=1}^{\bar{N}}\Big(\sqrt{\frac{1}{4}+4\,h(\lambda)^2\sin^2\frac{q_j}{2}}-\frac{1}{2}\Big).
\eea
The numbers $M$ and $\bar{N}$ figuring above is the total number of excitations in each SU(2), or equivalently, the number of $z_{1,k}$ and $\widetilde{z}_{2,k}$ oscillators. The letter $L$ in (\ref{bethe}) is the length of the spin chain and it can be expressed through the angular momentum $J$ and the excitation numbers as \cite{Hentschel:2007xn}
\bea
\label{Lrewrite}
L=J+\frac{1}{2}(M+\bar{N}).
\eea
Somewhat surprisingly (\ref{bethe}) is very similar to the corresponding set of equations in the AdS$_5 \times$ S$^5$ case \cite{Beisert:2005fw}. The only difference lies in the form of the interpolating function $h(\lambda)$ (which is constant in the AdS$_5$ case) and the phase factor. The phase factors in the two correspondences are related through \cite{Gromov:2008qe}
\bea
\label{phaserelation}
\sigma(p_k,p_j)_{AdS_5}=\sigma^2(p_k,p_j)_{CP_3}.
\eea
The Bethe equations (\ref{bethe}) are as they stand not very convenient for a large $P_+$ expansion since they are perturbative in both $P_+$ and $\lambda$ (or $\widetilde{\lambda}'$). We can put it in a form more appropriate if we rewrite the spin chain length, $L$, as
\bea
\label{Pplanguage}
J=\frac{1}{2}(P_++P_-), \qquad \lambda=\frac{P_+^2\,\widetilde{\lambda}'}{8\,\pi^2},
\eea
where we also expressed the original 't Hooft coupling in terms of the effective coupling defined in (\ref{scalingsandcoupling}). Expressing $L$ through the above and (\ref{Lrewrite}), together with the identity \cite{Beisert:2005fw}
\bea
\nn
\frac{\Phi(p_k)-\Phi(p_j)+i}{\Phi(p_k)-\Phi(p_j)-i}=\frac{x^+(p_k)-x^-(p_j)}{x^-(p_k)-x^+(p_j)}\cdot \frac{1-\big(x^+(p_k)\,x^-(p_j)\big)^{-1}}{1-\big(x^-(p_k)\,x^+(p_j)\big)^{-1}},
\eea
we rewrite (\ref{bethe}) as
\bea
\label{bethe1}
&&\Big(\frac{x^+(p_k)}{x^-(p_k)}\Big)^{\frac{1}{2}\,(P_++M+\bar{N})}=\\ \nn
&& \Big(\frac{x^+(p_k)}{x^-(p_k)}\Big)^{-\frac{1}{2}\,P_-}\prod_{k\neq j}^{M}\,\frac{x^+(p_k)-x^-(p_j)}{x^-(p_k)-x^+(p_j)}\cdot \frac{1-\big(x^+(p_k)\,x^-(p_j)\big)^{-1}}{1-\big(x^-(p_k)\,x^+(p_j)\big)^{-1}}\,\prod_{j=1}^{M}\sigma(p_k,p_j)\,\prod_{j=1}^{\bar{N}}\sigma(p_k,q_j), \\ \nn
&& \Big(\frac{x^+(q_k)}{x^-(q_k)}\Big)^{\frac{1}{2}\,(P_++M+\bar{N})}=\\ \nn
&& \Big(\frac{x^+(q_k)}{x^-(q_k)}\Big)^{-\frac{1}{2}\,P_-}\prod_{k\neq j}^{\bar{N}}\,\frac{x^+(q_k)-x^-(q_j)}{x^-(q_k)-x^+(q_j)}\cdot \frac{1-\big(x^+(q_k)\,x^-(q_j)\big)^{-1}}{1-\big(x^-(q_k)\,x^+(q_j)\big)^{-1}}\,\prod_{j=1}^{\bar{N}}\sigma(q_k,p_j)\,\prod_{j=1}^{M}\sigma(q_k,p_j),
\eea
At first glance this does not seem like a very useful reformulation of the original equations. However, using the ansatz
\bea
\label{momentumansatz}
p_k=\frac{p^0_k}{P_+}+\frac{p_k^1}{P_+^2}, \qquad q_j=\frac{q^0_j}{P_+}+\frac{q_j^1}{P_+^2},
\eea
it was shown in \cite{Hentschel:2007xn} that
\bea
\label{rewrite}
\Big(\frac{x^+(p_k)}{x^-(p_k)}\Big)^{-P_-}\, \prod_{k \neq i}^{K}\,\frac{1-\big(x^+(p_k)\,x^-(p_j)\big)^{-1}}{1-\big(x^-(p_k)\,x^+(p_j)\big)^{-1}}\, \prod_{j=1}^{K} \sigma^2(p_k,p_j)=1+\mathcal{O}(P_+^{-3}).
\eea
Since this is almost what appears in (\ref{bethe1}), we can eliminate the dependence on the scattering phase. Therefore, to order $P_+^{-2}$, we have
\bea
\label{bethe2}
&&\Big(\frac{x^+(p_k)}{x^-(p_k)}\Big)^{\frac{1}{2}\,(P_++M+\bar{N})}=\\ \nn
&& \prod_{k\neq j}^{M}\,\frac{x^+(p_k)-x^-(p_j)}{x^-(p_k)-x^+(p_j)}\cdot\Big( \frac{1-\big(x^+(p_k)\,x^-(p_j)\big)^{-1}}{1-\big(x^-(p_k)\,x^+(p_j)\big)^{-1}}\Big)^{\frac{1}{2}}\,\prod_{j=1}^{\bar{N}}\Big( \frac{1-\big(x^+(p_k)\,x^-(q_j)\big)^{-1}}{1-\big(x^-(p_k)\,x^+(q_j)\big)^{-1}}\Big)^{-\frac{1}{2}}\\ \nn
&& \Big(\frac{x^+(q_k)}{x^-(q_k)}\Big)^{\frac{1}{2}\,(P_++M+\bar{N})}=\\ \nn
&& \prod_{k\neq j}^{\bar{N}}\,\frac{x^+(q_k)-x^-(q_j)}{x^-(q_k)-x^+(q_j)}\cdot \Big(\frac{1-\big(x^+(q_k)\,x^-(q_j)\big)^{-1}}{1-\big(x^-(q_k)\,x^+(q_j)\big)^{-1}}\Big)^{\frac{1}{2}}
\,\prod_{j=1}^{M}\Big(\frac{1-\big(x^+(q_k)\,x^-(p_j)\big)^{-1}}{1-\big(x^-(q_k)\,x^+(p_j)\big)^{-1}}\Big)^{-\frac{1}{2}}.
\eea
What we gained from this is that for each order of $P_+$, these equations can be solved non perturbatively for $\widetilde{\lambda}'$. This was a feature which also was observed for the $AdS_5 / CFT_4$ case in \cite{Hentschel:2007xn}.

In the next section we will show that the energy shifts derived from the set of equations above exactly match the energies derived from the Hamiltonian (\ref{Hexp}).
\subsection{Large $P_+$ expansion}
Using the ansatz for the momentum (\ref{momentumansatz}), we can expand (\ref{bethe2}), which at leading order gives
\bea
\label{p0k}
p^0_k=4\,\pi \, m_k, \qquad q^0_j=4\,\pi \, \bar{n}_j,
\eea
where $m_k$ and $n_j$ takes values in the set of string mode numbers. For the next terms, $p^1_k$ and $q^1_j$, we get more complicated expression\footnote{We now simplify the notation using $\omega_{m_k}=\omega_k$. Which type of SU(2) excitation the indices takes values from should be clear from the context.}
\bea
\label{p1k}
&& p^1_k=-2\,\pi\,(M+\bar{N})\,m_k+16\,\pi\,m_k\Big\{\sum_{j\neq k}^M \frac{m_j(1+\omega_k+\omega_j)}{m_j(1+2\,\omega_k)-m_k(1+2\,\omega_j)} \\ \nn
&&+\sum_{j=1}^M\frac{m_j(m_k-m_j)\,\widetilde{\lambda}'}{(1+2\,\omega_k)(1+2\,\omega_j)-4\,m_k\,m_j\,\widetilde{\lambda}'}
-\sum_{j=1}^{\bar{N}}\frac{\bar{n}_j(m_k-\bar{n}_j)\,\widetilde{\lambda}'}{(1+2\,\omega_k)(1+2\,\omega_j)
-4\,m_k\,\bar{n}_j\,\widetilde{\lambda}'}\Big\},\\ \nn
&& q^1_k=-2\,\pi\,(M+\bar{N})\,\bar{n}_k+16\,\pi\,\bar{n}_k\Big\{\sum_{j\neq k}^{\bar{N}} \frac{\bar{n}_j(1+\omega_k+\omega_j)}{\bar{n}_j(1+2\,\omega_k)-\bar{n}_k(1+2\,\omega_j)} \\ \nn
&&+\sum_{j=1}^{\bar{N}}\frac{\bar{n}_j(\bar{n}_k-\bar{n}_j)\,\widetilde{\lambda}'}{(1+2\,\omega_k)(1+2\,\omega_j)-4\,\bar{n}_k\,\bar{n}_j\,\widetilde{\lambda}'}
-\sum_{j=1}^{M}\frac{m_j(\bar{n}_k-m_j)\,\widetilde{\lambda}'}{(1+2\,\omega_k)(1+2\,\omega_j)
-4\,\bar{n}_k\,m_j\,\widetilde{\lambda}'}\Big\}.
\eea
We want to use the solutions for $p^1_k$ and $q^1_j$ in the expression for the light-cone energy. To achieve this we expand $\Delta-J$ in (\ref{betheE})
\bea
&& \Delta-J= \\ \nn
&& \sum_{k=1}^{M}\Big(-\frac{1}{2}+\omega_k+\frac{1}{P_+}\,\frac{m_k \, p^1_k\,\widetilde{\lambda}'}{4\,\pi\,\omega_k}\Big)+ \sum_{k=1}^{\bar{N}}\Big(-\frac{1}{2}+\omega_k+\frac{1}{P_+}\,\frac{\bar{n}_k \, q^1_k\,\widetilde{\lambda}'}{4\,\pi\,\omega_k}\Big)+\mathcal{O}(P_+^{-3/2}),
\eea
and using the solutions for the rapidities gives the light-cone energy. This expression, which is presented in (\ref{bethefinalE}), is quite complicated and does not immediately resemble the solutions obtained from the string Hamiltonian in (\ref{finalE}). However, imposing the level matching constraint, together with some algebra, shows that the energy shifts obtained from the Bethe equations equal
\bea
&& \Delta E^{su2 \times su2}= \frac{\bar{N}\,M}{2\,P_+} -\frac{1}{4\, P_+}\Big\{\sum_{i=1}^M \sum_{j=1}^{\bar{N}}\frac{(m_i-\bar{n}_j)^2\,\widetilde{\lambda}'+2\, \omega_{m_i}\, \omega_{\bar{n}_j}}{\omega_{m_i}\, \omega_{\bar{n}_j}}\Big\} \\ \nn
&& -\frac{1}{16\, P_+}\Big\{\sum_{\substack{i,j \\ i \neq j}}^{\bar{N}} \frac{6\,\Omega_{\bar{n}_i+\bar{n}_j}^2-4\big(1+\omega_{\bar{n}_i}^2+\omega_{\bar{n}_j}^2\big)}{\omega_{\bar{n}_i}\, \omega_{\bar{n}_j}} +\sum_{\substack{i,j \\ i \neq j}}^M \frac{6\,\Omega_{m_i+m_j}^2-4\big(1+\omega_{m_i}^2+\omega_{m_j}^2\big)}{\omega_{m_i}\, \omega_{m_j}}\Big\}.
\eea
Which is identical to the energy shift from the string computation (\ref{finalE}).
\section{Summary and outlook}
In the present paper we have studied the near plane wave dynamics of a bosonic string propagating in an AdS$_4 \times \mathbbm{CP}_3$ background. Due to the recent proposal of \cite{Aharony:2008ug}, type IIA string theory in this background is supposedly equivalent to a three dimensional Chern Simons theory living on the boundary of the AdS space. This conjecture shares many similarities with the well studied $AdS_5 / CFT_4$ correspondence. In particular, it seems like many of the tools based on integrability are applicable also in this new proposal. Even though there has been a rapid progress in understanding the duality, nevertheless, it is safe to say that the integrable structures of the $AdS_4 / CFT_3$ correspondence still remains conjectural.

In the present paper we have added support for integrability in AdS$_4 \times \mathbbm{CP}_3$ by performing a direct comparison between string energies and predictions from a set of rewritten all loop Bethe equations (ULCB) \cite{Gromov:2008qe,Hentschel:2007xn}.

We started out with a detailed analysis of the cubic and quartic string Hamiltonian and its point particle dynamics. We removed the cubic terms with an unitary transformation and extracted the energy shifts for a certain subsector of the theory using first order perturbation theory.

We then calculated an exact all loop (in $\widetilde{\lambda}'$) expression for the energy shifts from the ULCB equations and successfully matched these with the energies obtained from the string computation. Since this is a result valid for an arbitrary number of string excitations, this calculation lends support for quantum string integrability.

There are several extensions of the current work. The most pressing is to make the model supersymmetric by adding fermions. Starting from \cite{Arutyunov:2008if}, this can be done along the lines of \cite{Frolov:2006cc}. As can be seen form the current paper, where the complications arising in the AdS$_4 \times \mathbbm{CP}_3$ background are brought to light, the addition of Fermions will be quite an involved calculation. Nevertheless, there should be no conceptual issues other than the ones described here, so deriving the full model should certainly be possible.

Another interesting line of research would be to investigate the role of the massive modes. As was discussed, $z_a$ and $y$ split up at the cubic level. However, it could be that they recombine if one interprets the cubic interactions correctly. For example, finding a suitable canonical transformation might shift the cubic part to quartic order in such a way that $y$ contracts with the AdS coordinates, restoring the SO(4) symmetry.

We plan to return to some of these questions in future works.
\acknowledgments
I would like to thank Gleb Arutyunov, Dmitri Bykov, Cecilia Flori, Sergey Frolov, Fabian Spill and Soo-Jong Rey for useful discussions and helpful comments. It is a pleasure to thank Tristan McLoughlin and Jan Plefka for an early collaboration on this project. I would especially like to thank Jan Plefka for suggesting the project. This work was supported by the International Max-Planck Research School for Geometric Analysis, Gravitation and String Theory.
\newpage
\appendix
\section{Generating functionals}
In this appendix we collect the various form of generating functionals and non-derivative interaction terms that was referred to in the main bulk of the text.
\subsection{Generating functional for point particle Hamiltonian}
Here we present the details of finding a generating functional $V$ that removes the interaction surviving the point particle limit.

Taking $\sigma \rightarrow 0$ in (\ref{Hexp}) removes all derivative terms, but leaves
\bea
\label{zeromodepiece}
&& \mathcal{H}_3^0=\\ \nn
&& \frac{1}{4 \sqrt{2\, P_+}}\Big\{\Big(x_i^2+4\,y^2+4\big(x_2 \, p_1-x_1\, p_2+x_4\, p_3-x_3\,p_4\big) -4\,p_i^2-8\,p_y^2\Big)y -4\,x_i\,p_y \, p_y,\Big\},
\eea
and
\bea
\label{zeromodepiece1}
&& \mathcal{H}_4^{,0}=\frac{1}{2\,P_+}\Big\{z_a^2\,(p_i^2+p_y^2)-p_a^2\,(y^2+\frac{1}{4}\,x_i^2)\Big\} \\ \nn
&& +\frac{1}{32 \, P_+}\Big\{4\,x_i^2\,y^2-(x_i^2)^2+24\,y^4+20\,x_i^2\, p_i^2+  12\big(x_1^2\,p_2^2+x_2^2\,p_1^2+x_3^2\,p_4^2+x_4^2\, p_3^2\big) \\ \nn
&& +4 \big((x_1^2+x_2^2)(p_3^2+p_4^2)+(x_3^2+x_4^2)(p_1^2+p_2^2)\big)+4\,x_i^2\,p_y^2+48\,y^2\,p_y^2 \\ \nn
&& +16\,y^2(x_2\,p_1-x_1\,p_2+x_4\,p_3-x_3\,p_4)+16\big((2\,x_1\,x_4-x_2\,x_3)p_1\,p_4+(2\,x_2\,x_3-x_1\,x_4)p_2\,p_3 \\ \nn
&& +(2\,x_2\,x_4+x_1\,x_3)p_2\,p_4+(2\,x_1\,x_3+x_2\,x_4)p_1\,p_3+x_1\,x_2\,p_1\,p_2+x_3\,x_4\,p_3\,p_4\big)+64\,y\,p_y\,x_i\,p_i\Big\}.
\eea
We want to construct a perturbative generating functional $V$, see (\ref{V}), with the property that
\bea
\{V, \mathcal{H}^0\}_{P.B}=-\mathcal{H}_3^0-\mathcal{H}_4^0+\mathcal{O}(P_+^{-3/2}).
\eea
One can easily see that a leading order term of $V$ as
\bea
\label{V3}
&& V_3=\frac{1}{\sqrt{2\,P_+}}\Big(p_i^2+\big(p_2\,x_1-p_1\,x_2+p_4\,x_3-p_3\,x_4\big)-\frac{1}{4}\,x_i^2-y^2\Big)\,p_y,
\eea
has the property, $\{V_3,\mathcal{H}_2^0\}=-\mathcal{H}_3^0$. Since this term starts at $\mathcal{O}(P_+^{-1/2})$, it also induces additional quartic terms
\bea
\label{quarticaddition2}
&& \mathcal{H}^0_{Add}=\frac{1}{2}\,\{V_3,\mathcal{H}_3^0\}=\\ \nn
&&-\frac{1}{2\,P_+}\Big\{p_y\,y\big(p_i\,x_i+4\,p_y\,y\big) +\frac{1}{2}\big(p_i^2+(p_2\,x_1-p_1\,x_2+p_4\,x_3-p_3\,x_4)-\frac{1}{4}\,x_i^2-y^2\big) \times \\ \nn
&&
\big(p_j^2+2\,p_y^2+(p_2\,x_1-p_1\,x_2+p_4\,x_3-p_3\,x_4)-\frac{1}{4}\,x_j^2-3\,y^2\big)
+ p_y^2\big(p_i^2+\frac{1}{4}\,x_i^2\big)\Big\}.
\eea
To remove these and $\mathcal{H}_4^0$, we can, after some trial and error, construct
\bea
\label{V4}
&& V_4= \\ \nn
&& \frac{1}{4\,P_+}\Big\{p_i\,x_i\big(y^2-p_j^2-p_y^2+p_a^2+z_a^2\big)+\frac{1}{4}\big(p_y\,y\,x_i^2+3\,p_i\,x_i\,x_j^2
-x_i^2\,p_a\,z_a\big) \\ \nn
&& -p_i^2\big(p_a\,z_a+3\,p_y\,y\big) -2\big(p_2\,x_1-p_1\,x_2+p_4\,x_3-p_3\,x_4\big)\big(p_y\,y+p_i\,x_i\big)+2\big(p_y\,y\,p_a^2-p_y^2\,p_a\,z_a\big)\Big\},
\eea
which has the desired property (\ref{V4prop}). With this we have managed to construct a generating functional that removes all non-derivative terms from the point particle Hamiltonian. However, note that this does not imply that the non-derivative terms can be ignored when $\sigma$ is non-zero.
\subsection{Unitary transformation}
For the unitary transformation that removes the cubic terms, the explicit form of (\ref{VpppVppm}) is
\bea
\label{Vunitaryrepresentation}
&& V^{+++}= \\ \nn
&& \frac{1}{16\,\sqrt{P_+}} \sum_{k,l,m}\delta_{k+l+m,0}\Big\{\frac{(\omega_k \, \omega_l \,\Omega_m)^{-1/2}}{\omega_k+\omega_l+\Omega_m}\Big(\big(1+4\,\widetilde{\lambda}'\,kl+4\,\omega_l(\omega_k+\Omega_m)\big)y_{-m}^\dagger \, x_{i,-k}^\dagger \,x_{i,-l}^\dagger \\ \nn
&& -4i(\omega_k-\omega_l)y_{-m}^\dagger \big( x_{2,-k}^\dagger \,x_{1,-l}^\dagger+x_{4,-k}^\dagger \,x_{3,-l}^\dagger\big)\Big) \\ \nn
&& +4\,\frac{(\Omega_k \, \Omega_l \,\Omega_m)^{-1/2}}{\Omega_k+\Omega_l+\Omega_m}\big(1-2\,\widetilde{\lambda}'\,kl+2\,\Omega_k \, \Omega_l\big)\,y_{-m}^\dagger \, y_{-k}^\dagger \,y_{-l}^\dagger\Big\},
\eea
and
\bea
\label{Vunitaryrepresentation1}
&& V^{++-}=\\ \nn
&& -\frac{1}{4\, \sqrt{P_+}}\sum_{k,l,m}\delta_{k+l+m,0}\Big\{\frac{(\omega_k \, \omega_l \,\Omega_m)^{-1/2}}{\omega_k-\omega_l+\Omega_m}\Big(\big((\omega_k-\omega_l)^2+\Omega_m(\omega_k-\omega_l)\big)y_{-m}^\dagger \, x_{i,-k}^\dagger\,x_{i,l} \\ \nn
&& -i(\omega_k+\omega_l)\,y_{-m}^\dagger \big( x_{2,-k}^\dagger \,x_{1,l}- x_{1,-k}^\dagger \,x_{2,l}+x_{4,-k}^\dagger \,x_{3,l}- x_{3,-k}^\dagger \,x_{4,l}\big)\Big) \\ \nn
\\ \nn
&& +\frac{(\Omega_k \, \Omega_l \,\Omega_m)^{-1/2}}{\Omega_k+\Omega_l-\Omega_m}\Big( 3-2\, \widetilde{\lambda}'\big(kl+km+lm\big)+2\big(\Omega_k\,\Omega_l-\Omega_k\,\Omega_m-\Omega_l\,\Omega_m\big)\Big)\,\,y_{-k}^\dagger \,y_{-l}^\dagger \,y_m\Big\} \\ \nn
&& -\frac{1}{8\,\sqrt{P_+}}\sum_{\substack{k,l,m\\ k \neq l \neq m \neq 0}}\delta_{k+l+m,0}\,\frac{(\omega_k \, \omega_l \,\Omega_m)^{-1/2}}{\omega_k+\omega_l-\Omega_m}\Big( \big((\omega_k+\omega_l)^2-2\,\Omega_m\,\omega_l\big) x_{i,-k}^\dagger \,x_{i,-l}^\dagger \,y_m \\ \nn
&&-2i(\omega_k-\omega_l)\big(x_{2,-k}^\dagger \,x_{1,-l}^\dagger +x_{4,-k}^\dagger \,x_{3,-l}^\dagger \big)y_m\Big).
\eea
Reality of the Hamiltonian demands,
\bea
V^{+--}=(V^{++-})^\dagger, \qquad V^{---}=(V^{+++})^\dagger.
\eea
In the last line of $V^{++-}$ we threw away the term where $k,l$ and $m$ are simultaneously zero, since for this term the denominator $\omega_k-\omega_l+\Omega_m$ is zero. Ignoring this contribution is allowed since the corresponding term in $G^{++-}$ is zero and does not contribute.

The above unitary transformation induces additional quartic terms, see (\ref{Hadd}). All terms that have an unequal number of creation and annihilation operators can be removed with further canonical transformations \cite{Frolov:2006cc}, so the relevant additional quartic terms are given by
\bea
\frac{i}{2} [V,\mathcal{H}_3]=i\big([V^{+++},G^{---}]+[V^{++-},G^{+--}]\big),
\eea
where we used that the additional part has to be Hermitian.
\section{Quartic Hamiltonian in oscillator expansion}
In this appendix we collect various expressions for the contributing parts of the original and the additional Hamiltonian. We start out with the original quartic contributions, which after putting the AdS$_4$ and the $y$ excitations to zero, equals
\bea
\label{H4orgOscillator}
&& \mathcal{H}_4^{CP_3}=-\frac{1}{16\,P_+}\,\sum_{\substack{klmn \\ k+l+m+n=0}}\,\big(\omega_k\,\omega_l\,\omega_m\,\omega_n\big)^{-1/2}
\Big\{\widetilde{z}_{2,-k}^\dagger\,z_{1,-l}^\dagger \, \widetilde{z}_{2,m}\, z_{1,n}\times \\ \nn
&& \Big(1-4\,(kl+mn)\widetilde{\lambda}'-8\,(km+ln)\widetilde{\lambda}'+6\,(\omega_k\,\omega_l+ \omega_m\,\omega_n)-2\,(\omega_k+\omega_l)(\omega_n+\omega_m)\Big) \\ \nn
&& +\frac{1}{2}\big(\widetilde{z}_{2,-k}^\dagger\,\widetilde{z}_{2,-l}^\dagger \, \widetilde{z}_{2,m}\,\widetilde{z}_{2,n}+z_{1,-k}^\dagger\,z_{1,-l}^\dagger \, z_{1,m}\,z_{1,n}\big)\times\Big(1-5\,(k+l)(m+n)\widetilde{\lambda}'+2\,(\omega_m\omega_n+\omega_k\,\omega_l) \\ \nn
&& -3\,(\omega_l\,\omega_n+\omega_k\,\omega_m)-5\,(\omega_l\,\omega_m+\omega_k\,\omega_n)\Big)\Big\}+\textrm{ non relevant terms.}
\eea
The non relevant terms are combinations like $z_{1,-k}^\dagger\,z_{1,-l}^\dagger \, \widetilde{z}_{2,m}\, \widetilde{z}_{2,n}$ which are present in the Hamiltonian (\ref{Hexp}) but nevertheless cancel among each other. From the worldsheet S-matrix point of view, this is quite obvious since processes like $Z_1\,Z_1 \Rightarrow Z_2 \, Z_2$ are not allowed due to charge conservation, see Table (\ref{tab:}).

For the additional term coming from the unitary transformation, we have for the first term, $[V^{+++},G^{---}]$, a contribution as
\bea
\label{HaddOsc1}
&& i\,[V^{+++},G^{---}]=\\ \nn
&& -\frac{1}{(32)^2\,P_+}\, \sum_{\substack{klmn \\ k+l+m+n=0}}\, \frac{(\widetilde{z}_{2,-k}^\dagger \,\widetilde{z}_{2,-l}^\dagger \, \widetilde{z}_{2,m} \, \widetilde{z}_{2,n}+z_{1,-k}^\dagger \, z_{1,-l}^\dagger \, z_{1,m}\, z_{1,n}) }{\Omega_{m+n}\,\sqrt{\omega_k\,\omega_l\,\omega_m\, \omega_n}}\times \frac{(\omega_k-\omega_l)(\omega_n-\omega_m)}{\omega_k+\omega_l+\Omega_{m+n}} \\ \nn
&& + \textrm{ non relevant terms}.
\eea
The $[V^{++-},G^{+--}]$ contribution is a little bit more complicated,
\bea
\label{HaddOsc2}
&& i\,[V^{++-},G^{+--}]= \\ \nn
&& \frac{1}{16\,P_+}\sum_{\substack{klmn \\ k+l+m+n=0}}\, \Big\{\frac{\widetilde{z}_{2,-k}^\dagger \, z_{1,-l}^\dagger \, \widetilde{z}_{2,m}\, z_{1,n}}{\sqrt{\omega_k \, \omega_l \, \omega_m \, \omega_n}}\Big[\Omega_{k+m}^{-1}\,\Big((\omega_k+\omega_m)(\omega_l-\omega_n) \\ \nn
&& +(\omega_k-\omega_m)(\omega_l-\omega_n)(\omega_m-\omega_k+\Omega_{k+m})+\frac{1}{\omega_l-\omega_n+\Omega_{k+m}}\big((\omega_k+\omega_m)(\omega_l+\omega_n) \\ \nn
&& +(\omega_k-\omega_m)(\omega_l+\omega_n)(\omega_m-\omega_k+\Omega_{k+m})\big)\Big)+\Omega_{l+n}^{-1}\,\Big((\omega_m-\omega_k)(\omega_l+\omega_n) \\ \nn
&& +(\omega_k-\omega_m)(\omega_l-\omega_n)(\omega_n-\omega_l+\Omega_{l+n})+\frac{1}{\omega_k-\omega_m+\Omega_{l+n}}\big((\omega_k+\omega_m)(\omega_l+\omega_n) \\ \nn
&& +(\omega_k+\omega_m)(\omega_n-\omega_l)(\omega_n-\omega_l+\Omega_{n+l})\big)\Big)\Big]+\frac{(\widetilde{z}_{2,-k}^\dagger \,\widetilde{z}_{2,-l}^\dagger \, \widetilde{z}_{2,m} \, \widetilde{z}_{2,n}+z_{1,-k}^\dagger \, z_{1,-l}^\dagger \, z_{1,m}\, z_{1,n}) }{4\,\Omega_{l+n}\,\sqrt{\omega_k\,\omega_l\,\omega_m\, \omega_n}}\Big[ \\ \nn
&& 4\,(\omega_m-\omega_k)(\omega_l+\omega_n)+4\,(\omega_k-\omega_m)(\omega_l-\omega_n)(\omega_n-\omega_l+\Omega_{l+n})\\ \nn
&& +\frac{1}{\omega_k-\omega_m+\Omega_{l+n}}\big((\omega_k+\omega_m)(\omega_l-\omega_n)(\omega_n-\omega_l+\Omega_{l+n})
-4\,(\omega_k+\omega_m)(\omega_l+\omega_n)\big) \\ \nn
&& +\frac{\Omega_{l+n}}{\Omega_{m+n}\,(\omega_k+\omega_l-\Omega_{m+n})}\,(\omega_k-\omega_l)(\omega_n-\omega_m)\Big]\Big\}+ \textrm{ non relevant terms}.
\eea
Using these two expressions allows us to calculate the additional energy shift corresponding to the unitary transformation.
\section{Expansion terms for the Bethe equations}
Using the solutions of the momentum components, (\ref{p0k}) and (\ref{p1k}), in (\ref{betheE}) gives
\bea
\label{bethefinalE}
&& \Delta E^{SU(2)\times SU(2)}= \\ \nn
&& \frac{\widetilde{\lambda}}{2\, P_+}\sum_{k=1}^M \Big\{-\frac{(M+\bar{N})\,m_k^2}{\omega_k} + \frac{8 \,m_k^2}{\omega_k}\Big(\sum_{j\neq k}^M \frac{m_j(1+\omega_k+\omega_j)}{m_j(1+2\,\omega_k)-m_k(1+2\,\omega_j)} \\ \nn
&&+\sum_{j=1}^M\frac{m_j(m_k-m_j)\,\widetilde{\lambda}'}{(1+2\,\omega_k)(1+2\,\omega_j)-4\,m_k\,m_j\,\widetilde{\lambda}'}
-\sum_{j=1}^{\bar{N}}\frac{\bar{n}_j(m_k-\bar{n}_j)\,\widetilde{\lambda}'}{(1+2\,\omega_k)(1+2\,\omega_j)
-4\,m_k\,\bar{n}_j\,\widetilde{\lambda}'}\Big)\Big\} \\ \nn
&& +\frac{\widetilde{\lambda}}{2\, P_+}\sum_{k=1}^{\bar{N}} \Big\{-\frac{(M+\bar{N})\,\bar{n}_k^2}{\omega_k} + \frac{8 \,\bar{n}_k^2}{\omega_k}\Big(\sum_{j\neq k}^{\bar{N}} \frac{\bar{n}_j(1+\omega_k+\omega_j)}{\bar{n}_j(1+2\,\omega_k)-\bar{n}_k(1+2\,\omega_j)} \\ \nn
&&+\sum_{j=1}^{\bar{N}}\frac{\bar{n}_j(\bar{n}_k-\bar{n}_j)\,\widetilde{\lambda}'}{(1+2\,\omega_k)(1+2\,\omega_j)-4\,\bar{n}_k\,\bar{n}_j\,\widetilde{\lambda}'}
-\sum_{j=1}^{M}\frac{m_j(\bar{n}_k-m_j)\,\widetilde{\lambda}'}{(1+2\,\omega_k)(1+2\,\omega_j)
-4\,\bar{n}_k\,m_j\,\widetilde{\lambda}'}\Big)\Big\}.
\eea
Showing that this equals the expression given by diagonalization of the string Hamiltonian in (\ref{finalE}) is a little bit involved. Easiest way to do this is to resort to Mathematica or some other computer program for algebraic manipulations\footnote{For people working with Mathematica, there is a very good package for quantum computations in \cite{cesar}.}. It is important to note though that expressions only equal upon imposing (\ref{momentumconstraint}).

\end{document}